\documentclass[journal]{IEEEtran}

\usepackage[noadjust]{cite}
\usepackage{graphicx}

\ifCLASSOPTIONcompsoc
  \usepackage[caption=false,font=normalsize,labelfont=sf,textfont=sf]{subfig}
\else
  \usepackage[caption=false,font=footnotesize]{subfig}
\fi

\usepackage{amssymb}
\usepackage{amsmath}
\usepackage{lscape}
\usepackage{multicol}
\usepackage{comment}
\usepackage{soul}
\usepackage[shortlabels]{enumitem}
\usepackage{multirow}
\usepackage{array}
\newcolumntype{P}[1]{>{\centering\arraybackslash}p{#1}}

\begin{document}

\title{Influence- and Interest-based Worker Recruitment in Crowdsourcing using Online Social Networks}

\author{Ahmed~Alagha,
        Shakti~Singh,~\IEEEmembership{Member, IEEE,}
        Hadi~Otrok,~\IEEEmembership{Senior Member, IEEE,}
        and~Rabeb~Mizouni
\thanks{All the authors are with the department of Electrical Engineering and Computer Science, Khalifa University, Abu Dhabi, UAE. S. Singh, H. Otrok, and R. Mizouni are also affiliated with the Center of Cyber Physical Systems (C2PS) in KU. (emails: ahmed.alagha@mail.concordia.ca; shakti.singh@ku.ac.ae; hadi.otrok@ku.ac.ae; rabeb.mizouni@ku.ac.ae).}

\vspace{-2em}}

\markboth{}%
{Alagha \MakeLowercase{\textit{et al.}}: Influence- and Interest-based Worker Recruitment in MCS using OSNs}

\maketitle

\begin{abstract}
Workers recruitment remains a significant issue in Mobile Crowdsourcing (MCS), where the aim is to recruit a group of workers that maximizes the expected Quality of Service (QoS). Current recruitment systems assume that a pre-defined pool of workers is available. However, this assumption is not always true, especially in cold-start situations, where a new MCS task has just been released. Additionally, studies show that up to 96\% of the available candidates are usually not willing to perform the assigned tasks. To tackle these issues, recent works use Online Social Networks (OSNs) and Influence Maximization (IM) to advertise about the desired MCS tasks through influencers, aiming to build larger pools. However, these works suffer from several limitations, such as 1) the lack of group-based selection methods when choosing influencers, 2) the lack of a well-defined worker recruitment process following IM, 3) and the non-dynamicity of the recruitment process, where the workers who refuse to perform the task are not substituted. In this paper, an Influence- and Interest-based Worker Recruitment System (\textit{IIWRS}), using OSNs, is proposed. The proposed system has two main components: 1) an MCS-, group-, and interest-based IM approach, using a Genetic Algorithm, to select a set of influencers from the network to advertise about the MCS tasks, and 2) a dynamic worker recruitment process which considers the social attributes of workers, and is able to substitute those who do not accept to perform the assigned tasks. Empirical studies are performed using real-life datasets, while comparing \textit{IIWRS} with existing benchmarks.
\end{abstract}

\begin{IEEEkeywords}
Crowdsourcing, Recruitment, Social Networks, Influence Maximization.
\end{IEEEkeywords}

\vspace{-1em}
\section{Introduction}
\label{Sec: Intro}
\IEEEPARstart{M}{obile} Crowdsourcing (MCS) is a distributed problem-solving model which utilizes the collective wisdom, ubiquity, and mobility of a large group of people carrying smart devices \cite{pedersen2013conceptual,abououf2020misbehaving}. The widespread use of smartphones and the recent advancements in information and communication technologies have brought MCS into the spotlight, capturing the attention of numerous applications and platforms, such as \textit{Waze}, \textit{Be My Eyes}, and \textit{Uber} \cite{wazewebsite,bemyewebsite,uberwebsite}.

Amongst the main challenges in MCS is the problem of recruiting workers who can provide high-quality services to meet the requirements of the task publisher \cite{azzam2016grs,AZZAM20181}. Here, the main challenge lies in maximizing the workers’ expected Quality of Service (QoS), while meeting the task requirements, mainly in terms of time and cost. Depending on the nature of the task, different criteria have been considered in the literature when assessing the QoS. Such criteria include the workers’ reputations, the coverage of the area of interest (AoI) provided by the workers, the residual energy in the workers’ devices, and the time needed to reach the task location \cite{abououf2018gale,azzam2016grs,AZZAM20181,SULIMAN20191158,ABOUOUF201952,alagha2022target,alagha2019data,alagha2020rfls,estrada2017crowd,xiong2015icrowd, alagha2021sdrs}. Furthermore, state-of-the-art mechanisms and frameworks have been proposed to optimize the QoS, using techniques such as Greedy Methods \cite{xiong2015icrowd}, Genetic Algorithm (GA) \cite{ABOUOUF201952,xiong2015icrowd,zhang20144w1h}, and Particle Swarm Optimization (PSO) \cite{estrada2017crowd}. 

\vspace{-1em}
\subsection{Problem Statement}
The recruitment of workers in MCS has been a major research problem over the past few years. The existing works related to recruitment in MCS commonly assume that a definite and pre-built pool of candidate workers, which is specific to the MCS task, is available. However, such an assumption is not realistic, especially in \textit{cold start} situations. The cold start refers to the case where a new MCS task has just been released, and an adequate pool of workers is yet to be built. For any MCS application, it takes time to accumulate an adequately large user group \cite{zhang20144w1h}. Additionally, even with an available pool of candidate workers, studies have shown that only a small portion of the registered workers, as small as 3.83\%, actually accept to perform an MCS task \cite{xu2018improving}. 

To overcome these shortcomings, few recent works have explored the idea of using Online Social Networks (OSNs), to provide more candidate workers and build larger pools for MCS systems \cite{wang2018social,xu2018improving,lu2019task, wang2020socialrecruiter}. Since OSNs (e.g. Twitter, Facebook, Instagram, etc.) play an important role in spreading information, opinions, rumors, etc., on a large scale, significant attention has been brought to the \textit{information diffusion} paradigm, which refers to the spread of information among interconnected nodes in the network. The idea is to first recruit influencers who spread the information about the desired tasks, aiming to attract more workers. This is achieved through \textit{Influence Maximization (IM)}, which aims to find the set of influencers that would maximize the influence, i.e. who would result in propagating the information to the maximum number of users. While these OSN-MCS works present contributions in laying the foundations of utilizing OSNs in MCS recruitment, they suffer from the following drawbacks:
\begin{enumerate}
    \item The proxy-based IM process in these works is done in an individual-based greedy manner, where candidates are ranked based on a metric, and those with the highest rank are iteratively added to the set of influencers. This does not guarantee the best set of influencers, since no group-based selection is performed, and later choices of influencers depend on the initial ones, which is one of the drawbacks of using greedy methods.
    \item There is a lack of an actual well-defined recruitment process following the IM process. Current OSN-MCS works assume that recruited workers are all those that have been influenced. This is inefficient since many MCS-related attributes, such as workers' exact GPS coordinates or the residual energy in their devices, cannot be obtained from the social network.
    \item Though the workers' interests are taken into consideration during the recruitment, yet the levels of interest are not considered. Individuals with high levels of interest in the task's domain are expected to yield better performance than those sharing the same interest at lower levels.
    \item There is a lack of a dynamic recruitment process, where workers that do not accept the task are substituted. 
    \item The recruitment is mainly designed for crowdsensing tasks, where workers do not need to travel to perform the task. This is unlike location-based crowdsourcing, where tasks, such as rating a restaurant or picking up a passenger, are in specific locations that the workers need to travel to. This requires special considerations to attributes like the distance to the task and the traveling time, which are missing from these works.
    \vspace{-1em}
\end{enumerate}
\subsection{Contribution}
\begin{figure}[!ht]
    \centering
    \includegraphics[width=0.4\columnwidth]{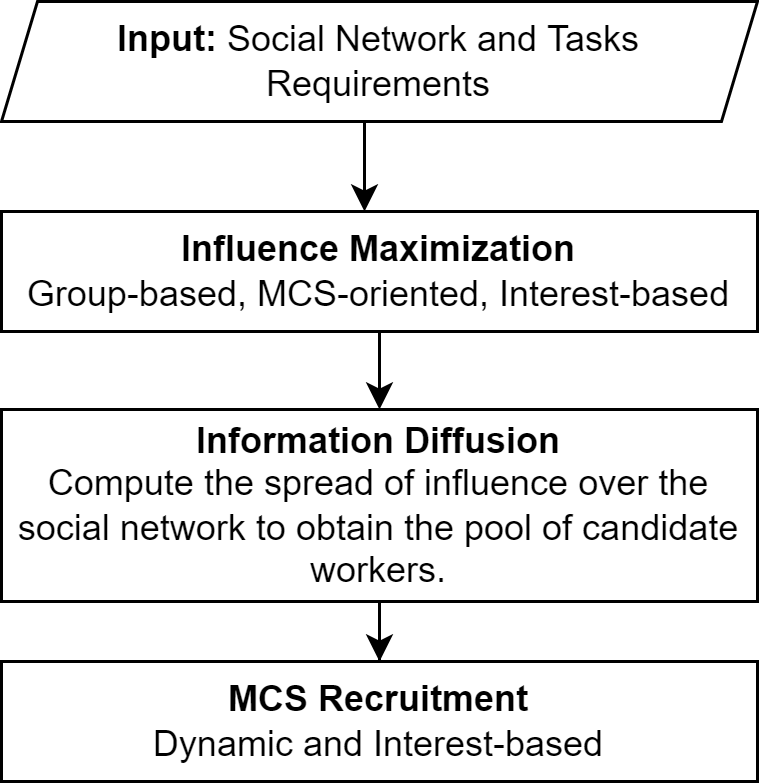}
    \caption{General overview of the proposed system (\textit{IIWRS}).}
    \label{Fig:IntroGeneralOverview}
\end{figure}

To tackle the aforementioned limitations, an Influence- and Interest-based Worker Recruitment System (\textit{IIWRS}) in MCS, using OSNs, is proposed. The system is composed of 3 stages. In the \textbf{first stage}, an MCS-, group-, and interest-based IM approach is designed to find the best set of influencers to advertise about a set of tasks, possibly from multiple domains. A group-based selection method is designed, using a GA, to choose the best set of influencers, while considering users' interests and MCS-related attributes. In the \textbf{second stage}, information diffusion models are used to compute the achieved influence given the selected set of influencers. The influenced users represent those who register in the MCS system as candidate workers, where more specific MCS-related attributes about them are collected. In the \textbf{third stage}, a dynamic and interest-based recruitment process is employed to recruit workers for a given task. This process is dynamic, i.e. workers who do not accept the task are substituted. It is also interest-based, where the workers' expected levels of interest are considered. The main aim of the full system is to obtain a large pool of interested candidate workers, while also ensuring dynamicity and resilience in the recruitment process to overcome the low rates of task acceptance. A general overview of the system is shown in Fig. \ref{Fig:IntroGeneralOverview}. The contributions of this work are:
\begin{enumerate}
    \item The design of a group-based IM approach to select a set of influencers by collectively assessing their rank, which is maximized using a GA. Individual attributes are designed and incorporated into group attributes, while ensuring balanced distribution between the group members.
    \item The design of an MCS- and interest-based IM approach. Influencers are selected based on a ranking metric, or a proxy, that considers users' interests in addition to MCS-related attributes, such as area coverage. 
    \item The design of a dynamic MCS workers recruitment process for location-based tasks that substitutes workers who do not accept to perform the assigned tasks.
    \item The design of an MCS workers recruitment process that considers the interest levels of the candidate workers, aiming to recruit workers with high levels of interests in the task's domain. As new users are brought through the social network, their social attributes (interest levels) are also obtained from the network, which are then used in the MCS recruitment process.
    \item The integration of proposed methods into one holistic system that uses OSN in assisting the recruitment process in MCS, with extensive testing and evaluation of the proposed system using real-life OSN and MCS datasets.
\end{enumerate}

To our knowledge, the proposed system is the first of its kind that integrates group-based selection in IM with a dynamic interest-based recruitment process, under one comprehensive solution. The evaluation of the proposed system is done using a real-life OSN obtained from Twitter, and real-life MCS datasets of individual's reputations and mobility patterns in the city of Cologne, Germany \cite{StackExchange,uppoor2011large,uppoor2013generation}.
\vspace{-1em}

\section{Related Works}
\label{Sec: RelatedWorks}
Since the aim of this work is to take advantage of the influence propagation in social networks, in order to enhance the workers’ recruitment in crowdsourcing systems, the related works section here is threefold: works related to MCS recruitment, IM, and OSN-assisted MCS recruitment. 
\vspace{-1em}
\subsection{Recruitment in MCS}
\label{subsec:Lit Recruitment in MCS}
Several research works have targeted the problem of workers recruitment in MCS systems. The common aim is to recruit the group of workers that maximizes the QoS while meeting the task requirements. The problem is usually modeled as a multi-objective optimization problem \cite{ma2020novel, alabbadi2021multi}, where a QoS is to be optimized using methods such Greedy algorithm \cite{xiong2015icrowd}, Genetic Algorithm \cite{azzam2016grs}, and Swarm Intelligence \cite{ma2020enhancing, estrada2017crowd}. Some of these works target crowdsensing tasks, where users submit reports without the need to travel to a certain location. The work in \cite{xiong2015icrowd} proposes a recruitment system that aims to maximize the spatial-temporal coverage of recruited workers under a fixed budget. It proposes a k-depth coverage metric and uses greedy methods to recruit workers in an individual-based manner, i.e. one by one. The works in \cite{azzam2016grs,AZZAM20181,alagha2019data,alagha2020rfls} use group-based recruitment systems, where candidate groups of workers are collectively assessed and the group with the best QoS is selected using genetic and/or greedy algorithms. The QoS depends on several attributes that vary between workers’ mobility patterns, area coverage, residual energy in workers’ devices, workers’ reputations, and the expected relevance of the data provided by them. Other works target crowdsourcing tasks, where tasks are location-based and workers have to travel to the task location in order to complete it. The works in \cite{abououf2018gale} and \cite{ABOUOUF201952} aim to match multiple tasks with multiple workers, using methods in clustering, GAs, and game theory, to maximize the QoS, which mainly use attributes like the time needed to reach the tasks, the reputation of the workers, and the confidence in the data they provide. A similar work in \cite{pu2016crowdlet} addresses the sequential worker recruitment problem, considering the worker's ability, arrival time, and cost of recruitment. The work in \cite{SULIMAN20191158} proposes a greedy-proof incentive mechanism, which considers the worker's contribution to the quality of information and the cost of performing the task. This helps outlying greedy workers, who request high costs for small tasks. Another work in \cite{jain2018quality} tackles the crowdsourcing (expertsourcing) problem using machine learning and game theory, where Multi-Armed Bandit (MAB) mechanisms are proposed select experts with aim of ensuring high accuracy levels, while incentivizing the experts to be truthful about their costs. The authors in \cite{zhang2021privacy} propose a Local Differential Privacy (LDP) mechanism, which helps in efficiently selecting a set of workers with a variety of attributes while providing privacy protection. The work in \cite{lu2021data} proposes a many-objective worker selection  mechanism based on an enhanced differential evolution algorithm (EDEA), which aims at selecting appropriate workers based on several objective functions. A work in \cite{abououf2021machine} proposes a recruitment system in MCS based on behavioral profiling, where machine learning is used to predict the probability of the workers performing a given task, based on their learned behavioral model.

The works above, make the assumption that a definite pool of workers is available. They also assume that once a task is pushed to workers, they will perform it, with no mechanisms that substitute a worker in case they do not perform the task. 
\vspace{-1em}
\subsection{Social Network and IM}
\label{Subsection: Lit SN and IM}
The problem of IM, which aims to find a subset of influential nodes in the network that can spread the information to the largest number of nodes, was found to be an NP-hard optimization problem \cite{liu2012time}. The classical IM method, adopted and extrapolated by many existing works, is based on a greedy framework, which selects a node as an influencer if it provides the maximum marginal gain to the influence of the current set \cite{li2018influence}. The influence is usually computed using information propagation models \cite{guille2013information}, which take an initial set of nodes (seed set: influencers) and estimate the influence achieved. Initial works, such as \cite{kempe2003maximizing,chen2013information}, use Monte Carlo (MC) simulations to estimate the influence of a seed set, and greedily select influential nodes if they provide the maximum marginal gain to this influence. Such an algorithm is proven to be prohibitively expensive, especially in large graphs, due to the high number of MC simulations required \cite{li2018influence}. 

Several works have emerged to tackle the IM problem in reduced complexity. The works in \cite{goyal2011celf++, zhou2015upper} proposes to estimate upper bounds of the influence spread in order to prune nodes with insignificant influence. Another approach is the state-of-the-art Reverse Influence Sampling (RIS) method \cite{borgs2014maximizing}, which is proven to tackle the IM problem in optimal time. RIS performs reverse sampling, by repeatedly starting at a random node $u$ and finding the reverse reachable (RR) set; i.e. the set of nodes that can reach $u$. The nodes that appear in the maximum number of (RR) sets are then selected as the influencers set. RIS has been extrapolated and improved in many subsequent works, where the complexity and memory consumption are further reduced, providing practical efficiency \cite{tang2014influence,wang2016bring, tang2015influence}. The work in \cite{guo2020influence} proposes an efficient random RR set generation algorithm using subset sampling, with the goal of reducing the computational cost. The recent work in \cite{he2019tifim} proposes a two-stage iterative framework for IM (TIFIM), which reduces the computation complexity by excluding less influential nodes. The first stage consists of an elimination mechanism, which excludes a large number of less influential nodes according to finite iterations of the First-Last Allocating Strategy (FLAS). The second stage uses Removal of the Apical Dominance (RAD) to determine seed nodes. 

Amongst the different types of IM algorithms are proxy-based IM algorithms. In such algorithms, a ranking metric (proxy) is used to estimate the influence. Simply, a ranking metric is used to assess candidate influencers, which are then chosen in a greedy manner, without the need to fully compute the influence. Proxy-based IM methods offer substantial performance improvements compared to MC-based approaches, and they are found to be practically efficient \cite{li2018influence}. Initial works proposed proxy-based methods using ranking metrics such as PageRank \cite{page1999pagerank} and Distance Centrality \cite{freeman1978centrality}. The work in \cite{narayanam2010shapley} proposes the use of Shapley value as a proxy, where nodes in the network are modeled as players in a coalitional game, and the diffusion process is captured as the coalition formation in the game. Another proxy algorithm, EaSyIM \cite{galhotra2016holistic}, is based on enumerating simple paths, and estimating the influence of each node by counting influence paths within a certain length. EaSyIM uses an iterative method that assembles Influence Ranking Influence Estimation algorithm (IRIE) \cite{jung2012irie} for global influence estimation, while accounting for the overlaps between different paths for better accuracy. The work in \cite{tang2017influence} proposes a hop-based algorithm, with theoretical guarantees, motivated by the fact that the majority of influence spread is produced within the first few hops of propagation. Recent works in \cite{keikha2018community,keikha2020influence} introduce the concept of network embedding, and propose deep learning techniques to learn the feature vectors of network nodes (network embedding) while preserving both local and global structural information. For each node, feature vectors that represent its neighborhood are formed, and cosine similarity is used to find the common nodes between vectors, which are then chosen as the influencers.

\vspace{-1em}
\subsection{Social Network-assisted MCS Recruitment}
\label{Subsection: Lit SN Recruitment}
The idea of utilizing OSNs to assist in MCS recruitment has been explored recently by few researchers. The works in \cite{wang2018social,lu2019task} propose proxy-based IM algorithms, in which ranking metrics are developed that consider the nodes’ degrees and the spatio-temporal coverage they provide. Influencers are then chosen in a greedy individual-based manner, based on the ranking metrics. A similar work in \cite{xu2018improving} considers the common social neighbors between two nodes using Jaccard similarity, in a ranking metric that is maximized using a greedy hill-climbing algorithm. Once influencers are selected, influenced users can participate through reverse auction, after which the platform selects a subset of users to perform the task. In \cite{wang2020socialrecruiter}, incentive mechanisms are proposed to encourage registered workers to propagate information about the task into the social networks by inviting friends. This is done in an optimization problem which maximizes the task completion, subject to constraints on the task propagation. 

As discussed earlier, although the works above introduced the idea of using social networks to assist with MCS recruitment, they have several drawbacks. Since the aim of IM is to maximize the collective influence of the seed set, individual- and greedy-based approaches do not yield the best results \cite{azzam2016grs}. Additionally, most of the works above are designed for crowdsensing tasks that are not location-specific, where users do not have to travel, and that is inapplicable to many crowdsourcing tasks. Moreover, it is always assumed that the exact locations (GPS coordinates) of the users can be obtained from the social network, which is not applicable in most social networks. It is also assumed that a user, once influenced, can directly be assigned a task. This is inefficient since MCS requires more information about the users and their devices (smartphones) that are not available on the social network. Finally, while some of these works aim to match the interests of the recruited workers with the domain of the task, they do not consider the interest levels which give an indication to the expected quality of reports submitted by the worker.
\vspace{-0.7em}

\section{Proposed System: \textit{IIWRS}}
\label{Sec: PropSystem}
In this section, an MCS Influence- and Interest-based Worker Recruitment System (\textit{IIWRS}), using OSNs, is presented. Given a set of MCS tasks $T=\{T_1,T_2,T_3,...\}$, the aim is to build a pool of candidate workers suited for the desired tasks. An OSN is used to propagate information about the tasks to the network’s users. Influenced users are those who register into the MCS platform, where more information about them and their devices are collected. For a given task $T_i$, the aim is to recruit a set of workers $W = \{W_1,W_2,W_3,...\}$ that would maximize the expected QoS and meet the task requirements.

A task $T_i$ in an MCS system is defined as a tuple in the form of $T_i=<L_i^T,{TC}_i^T,Rep_i^T,I_i^T>$, while a worker $W_j$ is defined as a tuple in the form $W_j= <L_j^W,Rep_j^W,{RE}_j^W,{SA}_j^W>$. An OSN is represented as a directed graph $G=(V,E)$, where $V$ is the set of nodes in $G$ (i.e. the OSN users) and $E$ is the set of directed edges. Each user $V_k$ is characterized by a tuple in the form of $V_k=<{ID}_k^V,L_k^V,I_k^V,{IC}_k^V,{OC}_k^V,P_k^V>$. Table \ref{Table: attributes} explains all the attributes defining $T_i$, $W_j$, and $V_k$. It is worth mentioning that $L_k^V$ is a general location since the exact GPS coordinates of a user cannot be obtained in most social networks; however many users give a general location in their profiles indicating the country, city, or area they live in. The user’s exact location ($L_j^W$) can be obtained later during the registration process. It is assumed here that the users’ interests are pre-mined.

\begin{table}[ht]
\caption{List of attributes, and their definitions, for the task, worker, and social network models.}
\setlength{\tabcolsep}{3pt}
\begin{center}
\begin{tabular}{P{37pt}|p{180pt}}
\hline
Attribute & Definition\\
\hline
$L_i^T$ & GPS coordinates of task $T_i$\\ 
$TC_i^T$ & Time constraint of task $T_i$\\
$Rep_i^T$	& Minimum reputation required by task $T_i$\\
$I_i^T$	& Interest/domain of task $T_i$\\
$L_j^W$	& GPS coordinates of worker $W_j$\\
$Rep_j^W$	& Reputation of worker $W_j$\\ 
${RE}_j^W$	& Residual energy in worker $W_j$’s device\\
${SA}_j^W$	& Set of worker $W_j$’s social attributes (connections, interests)\\ 
${ID}_k^V$	& ID number of user $V_k$\\
$L_k^V$	& General location of user $V_k$\\ 
$I_k^V$	& Set of interests of user $V_k$\\
${IC}_k^V$	& Set of incoming connections (followers) for user $V_k$\\
${OC}_k^V$	& Set of outgoing connections (followed by) for user $V_k$\\
$P_k^V$	& Set of posts by user $V_k$ related to the interests $I_k^V$\\

\hline
\end{tabular}

\end{center}
\label{Table: attributes}
\vspace{-1.5em}
\end{table}

\begin{figure*}
\centering
\includegraphics[width=0.66\textwidth]{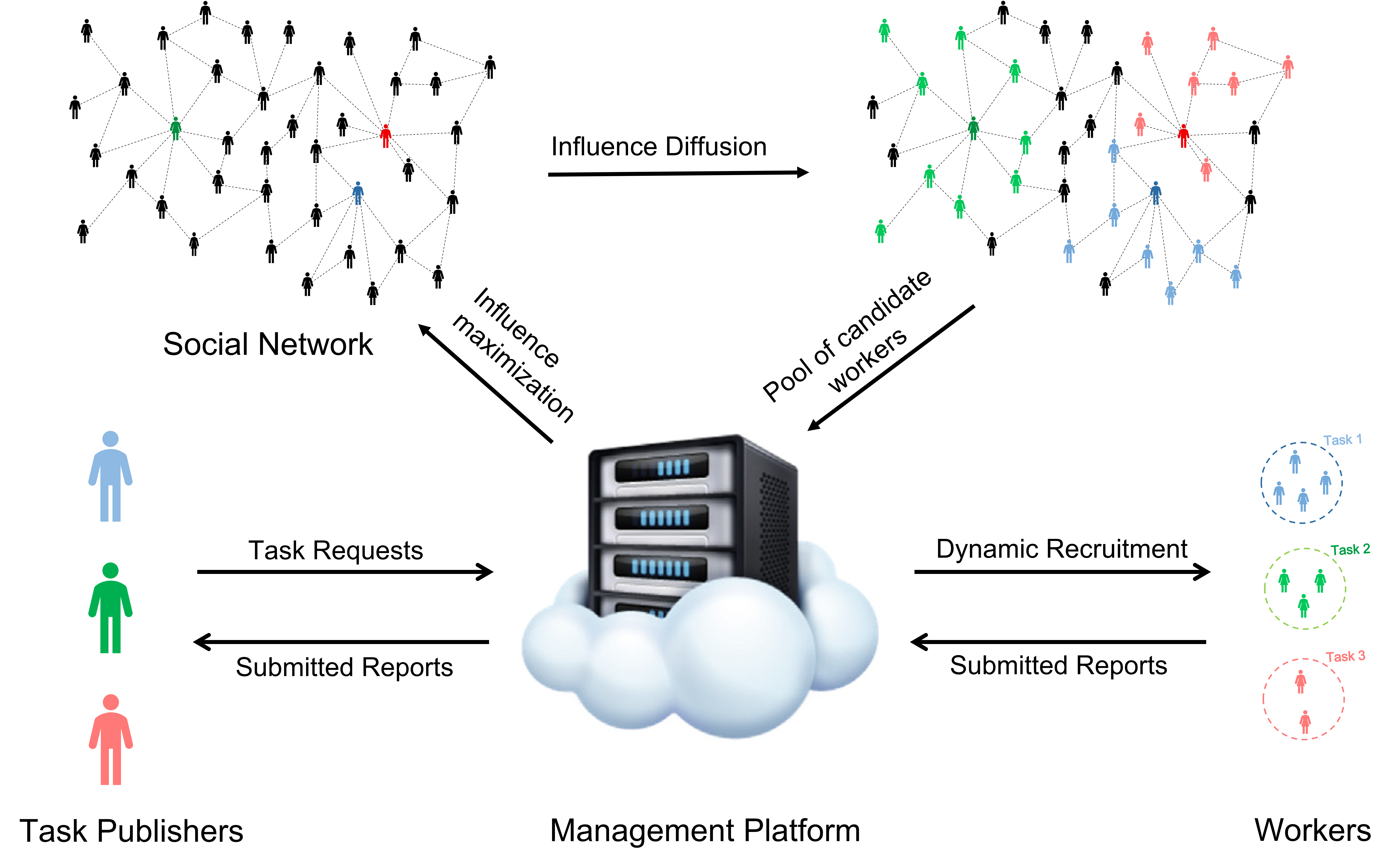}

\caption{An overview of the architecture of the proposed system. Given a set of tasks, submitted by task publishers, the management platform selects a set of influencers that maximizes the spread of influence to a targeted group in the network (here different colors indicate different interests). Influenced users are those who register in the system, representing the pool of candidate workers from which the management platform recruits workers for each task, and collects the submitted reports by the workers. Reports are assessed for quality assurance and sent to the task publishers.} 
\label{Figs: overall}
\vspace{-1em}
\end{figure*}

The proposed system, \textit{IIWRS}, is composed of 3 stages: 1) MCS-, group-, and interest-based IM (Section \ref{subsection: MCS&Interest IM}), 2) information diffusion (Section \ref{subsection: Info Diffusion and MCS Rec}), and 3) MCS dynamic recruitment (Section \ref{Subsec: MCS Recruitment}). The goal of the first stage is to find and employ influencers from the social network with specific characteristics that match the set of desired MCS tasks $T$, which could be from multiple domains. Employing influencers with interests that match these domains, and with locations near the desired tasks, is expected to propagate information about these tasks to a targeted pool in the network who are more suitable for such tasks, i.e. with similar interests and nearby locations. Once the best influencers are found, their influence is computed using an information diffusion model in the second stage, to obtain the set of influenced users. The influenced users are those who register in the MCS platform, where more information about them and their devices are collected. For a given task $T_i$, the recruitment process in the third stage recruits the best group of workers $W$, while considering their MCS- and interest-based attributes. The recruitment process is dynamic, where workers that do not accept task are substituted, which ensures that the quality of the task outcome is not affected. The architecture of the full system is illustrated in Fig. \ref{Figs: overall}.

\vspace{-1em}
\subsection{Stage 1: MCS-, Group-, and Interest-based IM}
\label{subsection: MCS&Interest IM}
In this stage, the aim is to select a set of influencers that would propagate the information about the MCS tasks to a targeted group in the network, with certain characteristics that match the desired tasks. A proxy-based IM is developed, where a ranking metric (proxy) is designed to find suitable influencers. Since it is desired to consider several attributes when choosing influencers, some of which are MCS-related (area coverage), proxy-based IM is favored over other approaches that only deal with nodes' connections, such as RIS. The following section motivates the use of a group-based selection methodology to choose the influencers, which performs better than individual-based methods.

\subsubsection{Motivation - Group-based selection} 
\begin{figure}[!ht]
    \centering
    \includegraphics[width=0.4\columnwidth]{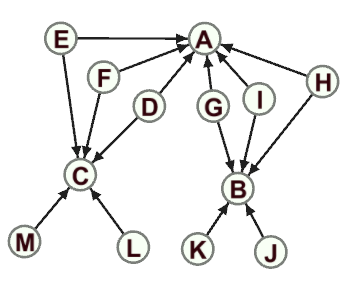}
    \caption{Example of a simple directed network.
    \label{Fig: MS2}}
    \vspace{-1.5em}
\end{figure}

The following case study displays the drawbacks in OSN proxy-based IM systems that are greedy and individual-based (similar to the system proposed in \cite{wang2018social}). As discussed in Section \ref{Sec: RelatedWorks}, proxy-based IM estimates the influence of candidate influencers using a ranking metric and selects influencers in a greedy manner. In this case study, a simple directed network, as shown in Fig. \ref{Fig: MS2}, is used. A directed network, like Twitter and Instagram, is one where an edge between two nodes has a specific direction, indicating that one node follows the other.
\vspace{-1em}

For simplicity, the ranking metric used here is the node’s \textit{in-degree}, which represents the number of incoming connections (followers). The aim is to find a set $S$ of 2 influencers that together give the highest number of unique followers. A greedy proxy-based IM process is as follows: The process starts with an empty set $S_0$, and at the first iteration, the node with the highest rank (in-degree) is added to the set. For each of the following iterations $t$, the node that brings the highest \textbf{marginal increase} to the in-degree is added to the set $S_t$. This is repeated until a set of the desired size is formed (2 in this case). Table \ref{Table: MS2} illustrates the process.
\vspace{-1em}
\begin{table}[ht]
\caption{Greedy IM process for selecting influencers.}
\setlength{\tabcolsep}{0pt}
\begin{center}
\begin{tabular}{|P{0.17\columnwidth}|P{0.15\columnwidth}|P{0.35\columnwidth}|P{0.15\columnwidth}|P{0.15\columnwidth}|}
\hline
Iteration ($t$) & $S_{t-1}$ & Candidate Nodes (Marginal In-degree) & $S_t$ & In-degree of $S_t$\\ 
\hline
1 & $S_0 =$ \{\} & A(6), B(5), C(5), D-M(0) & $S_1=$\{A\} & 6\\
\hline
2 & $S_1=$\{A\} & B(2), C(2), D-M(0) & $S_2=$\{A,B\} & 8\\
\hline
\end{tabular}

\end{center}
\label{Table: MS2}
\vspace{-1.5em}
\end{table}

The outcome of this process is a set $S=\{A,B\}$, with a total in-degree of 8, which indicates that the 2 influencers have 8 unique followers in total. However, when considering all the possible combinations of the influencers set, it can be found that $S=\{B,C\}$ yields an in-degree of $10$, which is higher than that found by the greedy method. This is mainly because a greedy algorithm lacks knowledge of what lies ahead of the current greedy state. This is applicable, not only while using the in-degree metric, but for any metric that relies on the group collectively. Additionally, while the best group was easy to find here by assessing all possible group combinations, this is impractical for larger networks due to the large number of combinations. For example, selecting $5$ influencers in a network of $10^4$ nodes requires exploring $8.3\times10^{17}$ different combinations. Hence, a scalable optimization method is needed to efficiently find the best group.

Since Proxy-based IM is proven to be practically efficient with lower computational complexity compared to MC-based IM \cite{li2018influence}, it is used in this work. It also allows for several attributes to be considered when assessing the influencers, and not only their connections. In proxy-based IM, a ranking metric is used to assess candidate influencers, then those with high ranks are selected. Since the aim here is to provide candidate workers to MCS tasks within certain areas and domains, some MCS-related attributes are taken into account, such as the coverage provided by the influencers and their interests. The ranking metric combines several group-based attributes, where each is based on the group members’ individual attributes, which are obtained from the social network. The following are the attributes considered in the ranking metric.

\subsubsection{Individual Attributes}
An individual attribute is one that characterizes a node in the network individually. In the proposed system, for a given node $V_k$, the following attributes are considered in the IM approach: 

\begin{itemize}[leftmargin=*]
    \item \textbf{Node Location ($L_k^V$)}: This attribute represents the general location of the node, i.e. user. Since the exact GPS coordinates of users in most social networks cannot be obtained, this attribute represents the general location of the node, that can be described as the country, city, or area of the user. In a practical application, this attribute could be mined from the user's profile or from their posts/activity. In this work, it is assumed that the general locations of users are available.
    \item \textbf{Node Interests ($I_k^V$)}: Mining users’ interests from OSNs has been an emerging research topic in recent years \cite{zarrinkalam2018mining,guy2013mining}. Practically, users' interests could mined from their profiles, their connections, and their activities. In this work, each user’s interest $I_k^V$ is assumed to be pre-mined and available as an input to the IM stage. The main goal is to select influencers with interests that match the domains of the MCS tasks, which is expected to yield a set of influenced nodes that also have similar interests.
    \item \textbf{Incoming Connections (${IC}_k^V$)}: This attribute represents the set of nodes following node $V_k$. The size of this set represents the number of followers a node has, which is significant in determining its potential influence.
\end{itemize}

\subsubsection{Group Attributes}
A group attribute collectively characterizes the entire group. Given the individual attributes of the group members, the group attributes considered in the IM process are as follows:
\begin{itemize}[leftmargin=*]
    \item \textbf{Group Distribution ($D^V(g)$)}: Given a certain AoI, $D^V(g)$ assesses how well the members of a group $g$ are distributed geographically, to give better coverage. Having influencers spread out would help reaching out to users from different areas, which would be later beneficial for more MCS tasks in varying locations. Given an AoI which is divided into $d$ subareas (e.g. cities or neighborhoods), the group distribution score, $D^V(g)$, is computed as follows:
    \vspace{-0.5em}
    \begin{equation}\label{eq: D(g)}
    \begin{split}
        D^V(g) = (c^g \times &\sum_{z=1}^{c^g} w^D_z) \times e^{-\sigma([\frac{m_1}{w^D_1}...\frac{m_z}{w^D_z}...\frac{m_{c^g}}{w^D_{c^g}}])}, \\  & \sum_{e=1}^{d} w^D_e = 1 , w^D_e > 0
    \end{split}
    \vspace{-0.5em}
    \end{equation}
    where $c^g$ is the number of subareas, out of $d$ subareas, covered by the group, $w^D_z$ is the weight (importance) given to each subarea, and $m_z$ is the number of group members located in subarea $z$. Practically, subareas could represent cities in a country (AoI), or neighborhoods in a city, and hence $D^V(g)$ would represent the portion of cities/neighborhoods covered by the group members. A subarea is considered to be covered if at least one member is located in it. Subareas are given weights depending on the problem. For example, in some scenarios, cities with higher populations (or larger areas) are expected to have high number of tasks, and hence would be given higher weights to give more importance to candidate influencers located in these areas. 
    
The first part of Equation \ref{eq: D(g)}, $c^g \times \sum_{z=1}^{c^g} w^D_z$, gives a higher score to groups that have greater coverage, i.e. cover more subareas, while considering the weights of these subareas. The second part of Equation \ref{eq: D(g)}, $e^{-\sigma([\frac{m_1}{w^D_1}...\frac{m_z}{w^D_z}...\frac{m_{c^g}}{w^D_{c^g}}])}$, uses standard deviation to ensure that group members are distributed in accordance with the weights. The maximum value of this part, i.e. $1$, is guaranteed only if $\frac{m_1}{w^D_1} = \frac{m_2}{w^D_2}= ... = \frac{m_{c^g}}{w^D_{c^g}}$. Otherwise, the score is reduced depending on how different the values are. For example, given $3$ subareas with the weights [$w^D_1=0.2$, $w^D_2= 0.2$, $w^D_3= 0.6$], a group with the distribution [$m_1=1$,$m_2=1$,$m_3=3$] would yield a standard deviation of $0$, which in turn yields $e^0=1$. This is a desired outcome that represents fair distribution, since subarea $3$ is three times more important than the other subareas, and hence it has thrice the number of members.  
    
    \item\textbf{Group Interests ($I^V(g)$)}: A set of MCS tasks could be under one or multiple domains, such as environmental monitoring, transportation, gaming, music, etc. The final pool of candidate workers, to be built from the social network, should ideally share interests that match the domains of the required MCS tasks. One way to achieve this is to ensure that the chosen influencers have these required interests. Given a set of interests related to the desired MCS tasks, $I^T=[I_1^T,I_2^T,...,I_{NI}^T]$, where $NI$ is the number of interests, the aim is to prioritize candidate influencers having these interests. Hence, for a set of candidate influencers, the $I^V(g)$ parameter is computed as:
    \vspace{-0.5em}
    \begin{equation}\label{eq: I(g)}
    \begin{split}
        I^V(g) = (\sum_{x=1}^{NI}& w_x^IE_x) \times e^{-\sigma([\frac{E_1}{w^I_1}...\frac{E_x}{w^I_x}...\frac{E_{NI}}{w^I_{NI}}])}, \\  &\sum_{y=1}^{NI} w^I_y = 1, w^I_y > 0 
    \end{split}
    \end{equation}
    where $E_x$ is the number of group members having interest $I_x^T$ and $w^I_x$ is the weight given to this interest. Similar to Equation \ref{eq: D(g)}, the first part of Equation \ref{eq: I(g)}, $\sum_{i=x}^{NI} w^I_xE_x$, gives higher scores to groups with larger number of members per interest, taking into consideration the interests’ weights. The second part of the equation, $e^{-\sigma([\frac{E_1}{w^I_1}...\frac{E_x}{w^I_x}...\frac{E_{NI}}{w^I_{NI}}])}$, uses the standard deviation to reduce the score in case the ratios $\frac{E_x}{w^I_x}$  are not equal, which indicates uneven distribution of members per weighted interests.
    \item \textbf{Group Unique Followers ($U^V(g)$)}: The main goal of the IM stage is to spread the information about the MCS tasks to as many interested and suitable users as possible. One of the main indications for the influence of a user is the number of followers it has. The parameter $U^V(g)$ considers the in-degree, i.e. the number of followers $IC_k^V$, of the group members, while eliminating any duplicates. In other words, the aim is to maximize the unique number of followers.
    \begin{equation}\label{eq: U(g)}
        U^V(g) = unique\_followers(g)
    \end{equation}
\end{itemize}

\subsubsection{Optimization Problem Definition}
\label{subsubsection: Optimization problem Def}
To assess a candidate group of influencers, the group is given a score based on the following ranking metric:
    \begin{equation}\label{eq: R(g)}
        R(g) = \sqrt[3]{D^V(g) \times I^V(g) \times U^V(g)}
    \end{equation}
The geometric mean is used to compute $R(g)$, since the parameters represent different attributes that could be of different ranges, and the geometric mean normalizes these ranges such that no single range dominates \cite{enyi2019analyzing}. The geometric mean has been commonly used to combine attributes of different ranges in MCS recruitment works \cite{alagha2020rfls, alagha2021sdrs}. Any alternative method, such as the product of the given attributes, could be used equivalently, as long as it normalizes the attribute ranges.

An individual node, to be considered in a candidate group, must not violate any of the following constraints:
\begin{itemize}[leftmargin=*]
    \item 	The location of each node is within the AoI ($L_k^V \in  AoI$).
    \item 	The number of incoming connections is higher than the minimum required degree ($size(IC_k^V) \geq MinDegree$).
    \item The node should have at least one interest that is common with the set of required tasks $T$.
\end{itemize}
The $MinDegree$ is a threshold which determines the minimum in-degree required for a node to be considered in a candidate set of influencers. The bigger the network, the higher the threshold should be. Generally, a large portion of any social network is composed of nodes with very low in-degree. Such nodes should not be considered as potential influencers to save computation time.

In addition to the individual constraints implied on each candidate node, a candidate group $g$ must not violate the following constraint:
\begin{itemize}
    \item 	Each interest in $I^T$ is satisfied by at least one member.   
\end{itemize}

\subsubsection{Proposed IM Approach}
As discussed in Section \ref{subsection: MCS&Interest IM}, the proposed IM approach relies on a group-based selection, where the aim is to maximize $R(g)$, given a set of tasks $T$. Given the current size of social networks, a large number of group combinations are possible, and it is computationally exhaustive to assess all of them and choose the best group. As mentioned earlier, a case of selecting $5$ influencers in a network of $10^4$ nodes requires exploring $8.3\times10^{17}$ different combinations. Instead, a GA \cite{azzam2016grs} is used to find the group that maximizes $R(g)$. GA is favored over other heuristics because: (i) GA is efficient and scalable in searching for the optimal solution in large search spaces \cite{madni2016appraisal}, (ii) GA is experimentally proven to be computationally faster than many other heuristics like differential evolution (DE) and particle swarm optimization (PSO) \cite{lim2013performance}, and (iii) the efficiency of GA in optimizing group-based selection has been reported in several existing works \cite{azzam2016grs,AZZAM20181,ABOUOUF201952,alagha2019data}. 

\begin{figure}[!ht]
    \centering
    \includegraphics[width=0.7\columnwidth]{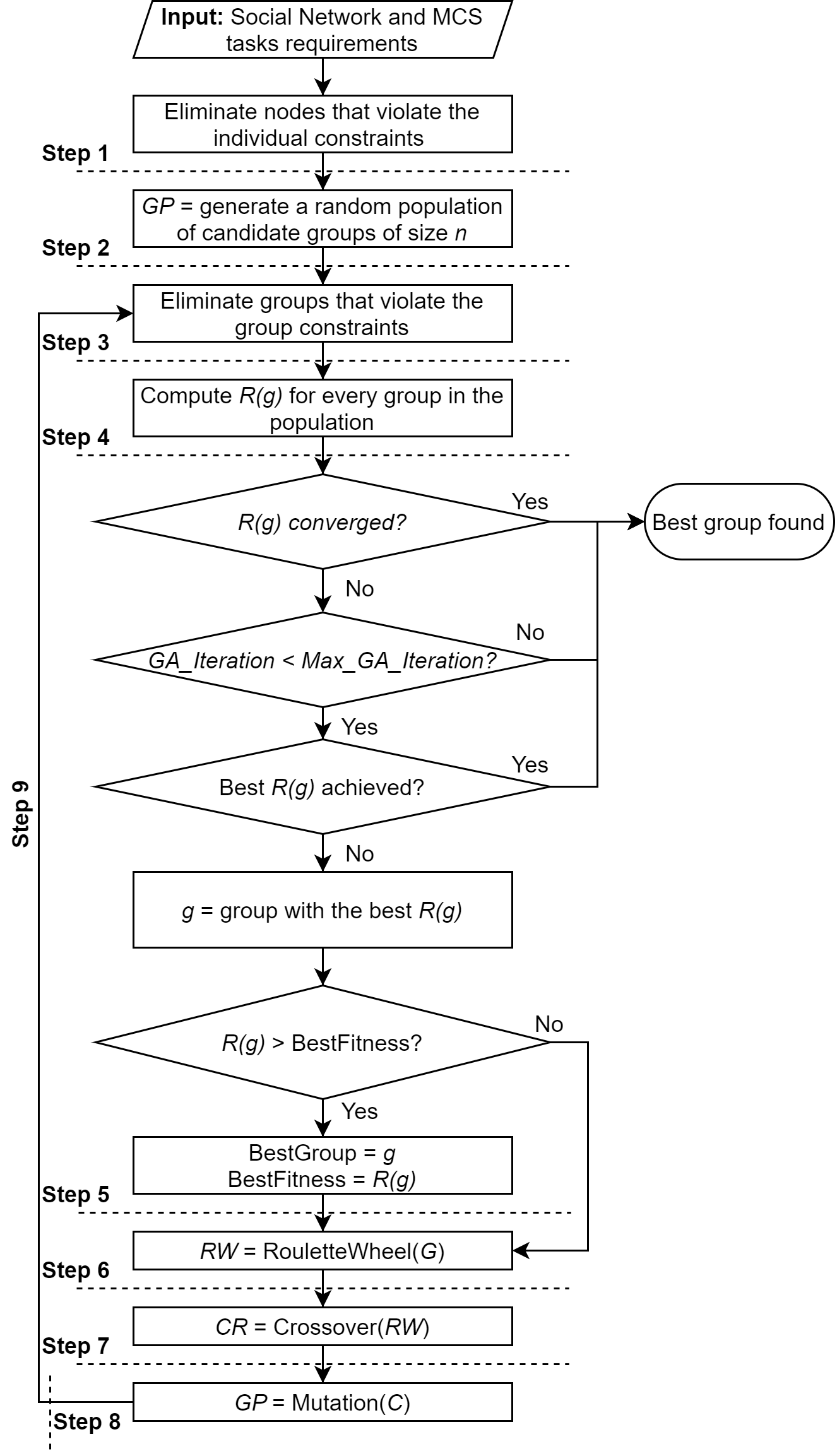}
    \caption{Flowchart of the MCS-, Group-, and Interest-based IM process.}
    \label{Fig:IM Process Flowchart}
    \vspace{-1.5em}
\end{figure}

A detailed description of the IM process is shown in Fig. \ref{Fig:IM Process Flowchart}. It is worth mentioning that the process terminates in Step 5, i.e. the best group is found, if one of 3 conditions is met: 1) $R(g)$ converges, which means that the best value of $R(g)$ does not improve for a certain number of GA iterations, 2) the maximum number of GA iterations is reached, or 3) the maximum possible score for $R(g)$ is achieved.

\vspace{-1em}
\subsection{Stage 2: Information Diffusion}
\label{subsection: Info Diffusion and MCS Rec}
In a real-life application, recruited influencers spread (i.e. diffuse) information about the MCS tasks through posts on the OSN. Users influenced by such posts are those who register in the system to be part of the candidate pool of workers. To simulate this entire process, this work uses an  \textit{information diffusion} model \cite{li2018influence}. Information diffusion helps in obtaining an \textbf{\textit{estimate}} of what the influence would be (i.e. the number of registered workers) if the model is implemented in real-life. Information diffusion models compute the spread of influence starting from a seed set of influencers. Given a social network $G$, a seed set $S \in V$, and a diffusion model, the influence spread $\sigma(S)$ is the expected influence achieved by $S$. 

A generic diffusion model associates a node $u \in V$ with a status of either \textit{active} (influenced) or \textit{inactive} (not influenced). Initially, the nodes in the seed set $S$ are active, while the rest of the nodes in $V$ are inactive. The active nodes can influence and activate their neighbors, which in turn can also activate their neighbors, and so on. The process of diffusion terminates when no more nodes can be activated.

One of the most commonly used diffusion models is the Independent Cascade (IC) model \cite{goldenberg2001talk}. In this model, a node has a chance of being activated by each of its active neighbors independently. Each directed edge connecting two nodes, $a$ and $b$, is associated with an activation probability $p_{a,b}$, which reflects the influence $a$ has on $b$. Let the set of active nodes be denoted as $A$, and the set of neighbors of a given node $a$ be denoted as $NB(a)$. Following the IC model, the diffusion process occurs in discrete steps and is given as follows: at step $t$, each active node $a \in A$ will attempt to activate each of its inactive neighbors $q \in NB(a)$, based on a probability $p_{a,b}$. An active node $a$ has only one chance to activate its neighbors, after which it stays active and stops activation attempts. The process repeats, where newly activated nodes attempt once to activate their neighbors, until no more nodes can be activated, and hence the diffusion process terminates, and the result is a set of active nodes $A(S)$. The IC model is a progressive one, i.e. it assumes that an activated node cannot be deactivated later. To determine $p_{a,b}$, a common method is to use existing data of user interactions (such as replies, forwards, likes) \cite{li2018influence}. Hence, based on data from online social networks, it is assumed here that the influence probability is known, with a value of $p_{a,b}=0.02$, which reflects the engagement rate an average influencer has \cite{statista1,statista2}.

In a real-life application, influenced users (i.e. those who register in the MCS system) provide more information regarding their MCS attributes, such as exact GPS coordinates and residual energy in their device, once they register in the system. To simulate this, once the influence is computed in simulations, each influenced user is assigned a set of MCS attributes, which are either generated randomly or obtained from existing MCS datasets, as discussed later in Section \ref{Subsec: Dataset}. The set of influenced users is thereafter referred to as the pool of candidate users.

\vspace{-1em}
\subsection{Stage 3: MCS Recruitment}
\label{Subsec: MCS Recruitment}
Following the IM and information diffusion stages, a pool of candidate workers is ready for the recruitment stage. For a given task $T_i$, characterized by $T_i=<L_i^T,{TC}_i^T,Rep_i^T,I_i^T>$, the aim is to recruit a group of workers $W=\{W_1,W_2,W_3,...\}$ that would maximize the expected QoS and meet the task requirements. Since each task belongs to a specific domain, it is desired to recruit workers who, not only share interests related to this domain, but also show high levels of these interests. For a resilient system with high QoS, it is also important to ensure the completion of the desired task, hence it is crucial to guarantee that all recruited workers accept to perform the task.

\subsubsection{Recruitment Parameters}
\label{subsubsection: Rec Parameters}
There are several attributes considered when recruiting a worker. As presented previously, a worker $W_j$ is characterized by $W_j=<L_j^W,Rep_j^W,{RE}_j^W,{SA}_j^W>$. These attributes are used in the following recruitment parameters.

\begin{itemize}[leftmargin=*]
    \item \textbf{Traveling Time (${Tr}_j^W$)}: Since the proposed system targets location-based tasks, the ${Tr}_j^W$ parameter determines the expected traveling time needed for the worker to reach the task. ${Tr}_j^W$ is calculated as the distance between the worker’s location $L_j^W$ and the task location $L_i^T$ over the average speed of the worker. The average speed of the worker can be estimated, assuming the traffic conditions are known.

    \item \textbf{Interest Level (${IL}_j^W$)}: One of the main goals of the proposed system is to target a group of workers that have interests matching the domains of the desired tasks. However, choosing workers with the desired interest is not sufficient if the interest levels are not considered. For example, a user who tweets on a daily basis about food and restaurants is expected to be more committed and informative regarding a task in this domain, compared to a user who occasionally tweets about this topic.  In this work, the interest level ${IL}_j^W$ of a worker is obtained by analyzing their social attributes ${SA}_j^W$, such as their posts and the users they follow. Hence, the interest level ${IL}_j^W$ score for a worker is given as:
    \begin{equation}
        {IL}_j^W=avg(P_j,F_j)
        \label{eq: IL}
    \end{equation}
    where $P_j$ is the number of posts per unit time by $W_j$ related to the interest matching the task's domain, and $F_j$ is the number of users followed by $W_j$ who share the same interest. Both $P_j$ and $F_j$ scores are normalized over the maximum value among candidate workers in the pool.
    
    \item \textbf{Residual Energy (${RE}_j^W$)}: The residual energy parameter, which indicates the battery level in the user’s smart devices, is a classical parameter considered by many MCS recruitment works \cite{azzam2016grs,AZZAM20181,ABOUOUF201952,alagha2019data}. It may also reflect the confidence in a worker to attempt and complete a task \cite{ABOUOUF201952}.

    \item \textbf{Reputation ($Rep_j^W$)}: The worker’s reputation reflects their historical performance. It is computed as the ratio of correctly submitted reports to all submitted reports \cite{azzam2016grs,AZZAM20181,ABOUOUF201952}. Users who register for the first time are given a reputation of $0.5$, which then gets updated based on their performance

\end{itemize}

\subsubsection{Optimization Problem Definition}
\label{subsubsection: recruitment optimization problem}
Since the above MCS recruitment parameters are individual-based, hence the expected QoS metric used to assess a candidate worker $W_j$ is given as follows:
\begin{equation}
    {QoS}_j^W= \sqrt[4]{{RE}_j^W \times {IL}_j^W \times {\tau}_j^W \times Rep_j^W}
    \label{eq: recruitment QoS}
\end{equation}
where ${\tau}_j^W$ is a decreasing function with respect to ${Tr}_j^W$; i.e. it decreases as ${Tr}_j^W$ goes higher. It is computed as follows \cite{estrada2017crowd, yu2015quality}:
\vspace{-0.5em}
\begin{equation}
    \label{eq: tau}
    {\tau}_j^W= 1-max(0,min[log_{{TC}_i^T}({Tr_j^W}),1]) 
\end{equation}
The time constraint ${TC}_i^T$ is used as the base for the log operation to ensure that any value for $Tr$ beyond the time constraint results in $\tau = 0$, since the worker in this case cannot reach the task location in time. The geometric mean is used for the QoS equation for reasons similar to those discussed in Section \ref{subsubsection: Optimization problem Def}. A worker $W_j$ must satisfy the following constraints to be considered in the recruitment process:
\begin{itemize}[leftmargin=*]
    \item $I_i^T \in I_j^W$: the task's domain matches one of the worker’s interests.
    \item ${Tr}_j^W \leq {TC}_i^T$: the worker is able to travel to the task within the time constraint.
    \item ${QoS}_j^W \geq {QoS}_{min}$, where ${QoS}_{min}$ is a minimum QoS threshold specified by the task publisher.
\end{itemize}

\subsubsection{Proposed Recruitment Process}
\begin{figure}[!ht]
    \centering
    \includegraphics[width=0.5\columnwidth]{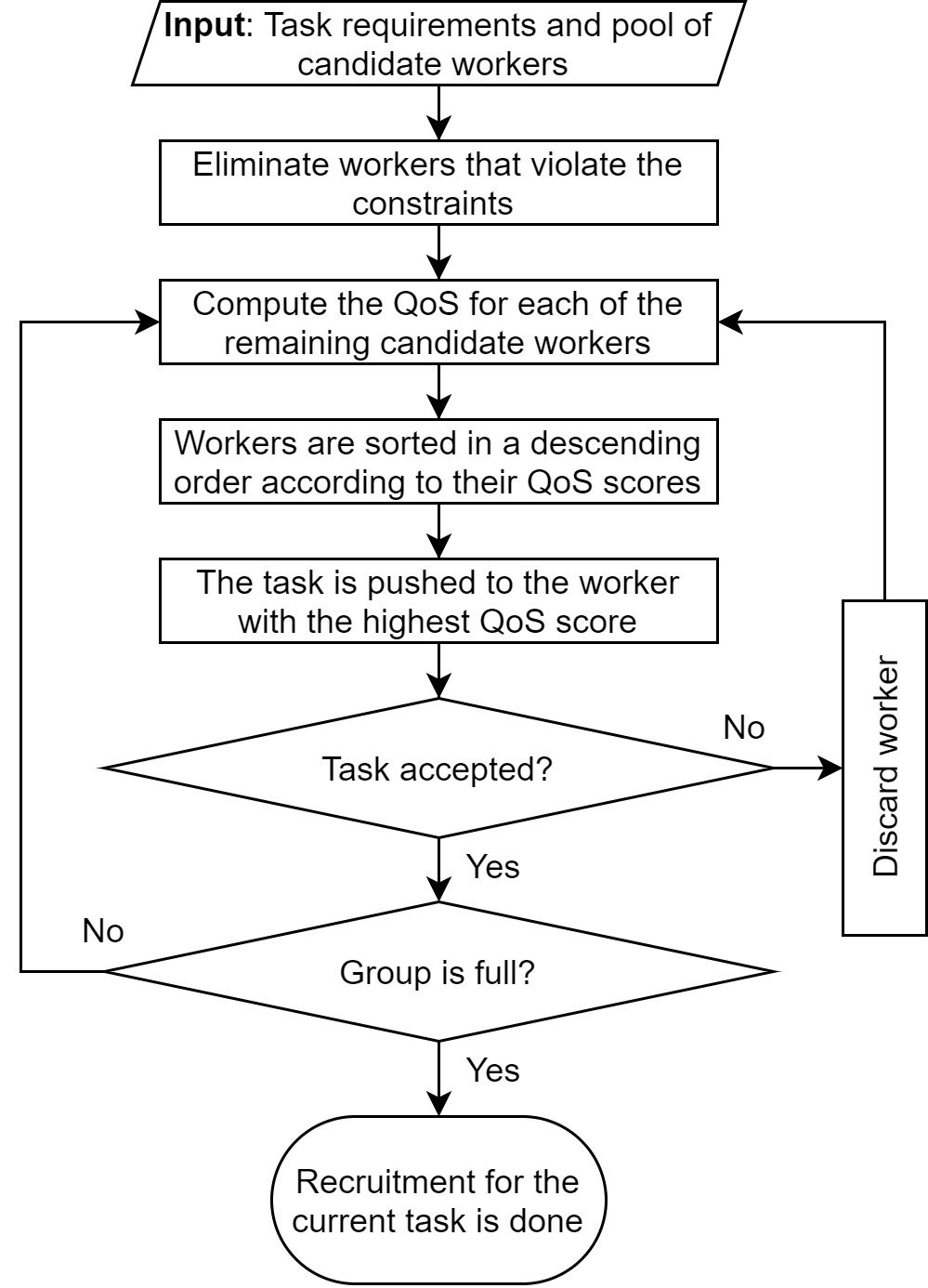}
    \caption{Flowchart of the recruitment approach.}
    \label{Fig:recruitment Flowchart}
    \vspace{-1em}
\end{figure}

Given a task $T_i$ and the pool of candidate workers, the recruitment process aims to recruit a group of size $gs$ that maximizes the expected QoS, while meeting the task requirements. Unlike the IM process, where collective assessment of the possible sets of influencers was needed, and hence a GA group-based method was used, workers here perform the tasks individually. Hence, an individual-based greedy method is used. Fig. \ref{Fig:recruitment Flowchart} describes the proposed recruitment approach.

\section{Simulation and Evaluation}
\label{sec: simulations}
This section reports and discusses the performance of the proposed system, \textit{IIWRS}, using real-life datasets, and compared to some existing systems \cite{wang2018social,lu2019task, azzam2016grs, page1999pagerank}.
\subsection{Dataset Description}
\label{Subsec: Dataset}
\begin{figure}[!ht]
    \centering
    \includegraphics[width=0.4\columnwidth]{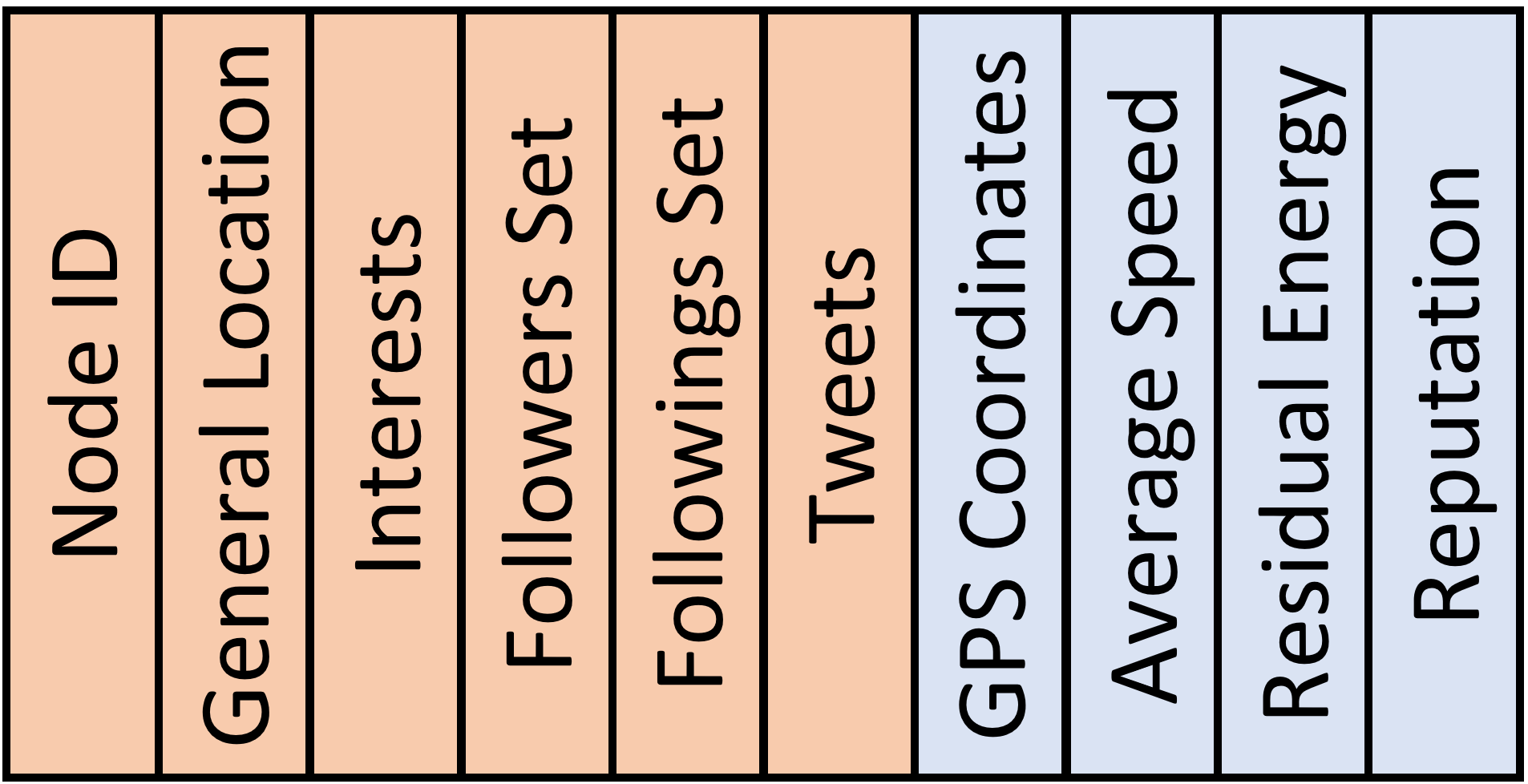}
    \caption{The structure of the dataset. The first 6 attributes are obtained directly from the OSN, while the rest are obtained during the MCS registration process.}
    \label{Fig: dataset structure}
    \vspace{-0.7em}
\end{figure}

To evaluate the proposed system, the dataset needs to describe the nodes in an OSN, where each node $V_k$ is characterized by: 1) a unique ID, 2) general location, 3) interests, 4) set of followers, 5) set of nodes it is following, and 6) its posts/tweets. Given that a user registers in the MCS system as a candidate worker, the dataset also needs to describe 7) their exact GPS coordinates, 8) their average speed, 9) the current residual energy in their device, and 10) their reputation. The structure of the full dataset is shown in Fig. \ref{Fig: dataset structure}. 

The OSN dataset used in this work is extracted from Twitter using Twitter API\footnote{https://developer.twitter.com}. This network is used to extract the first 6 attributes shown in Fig. \ref{Fig: dataset structure}. For this work, the network is reduced to one with 5 interests: \textit{sports}, \textit{music}, \textit{movies}, \textit{books}, and \textit{gaming}. More than 55 hashtags and keywords were used to extract tweets related to these interests, for the period from 3rd-10th of July 2020. The dataset considers only geotagged tweets located in the UK. The user profiles for these tweets are extracted, along with information such as their IDs, followers list, following list, and general locations. Each user is assigned a set of interests based on the extracted tweets. For example, if a user has tweeted about sports and music, then these topics are added to their list of interests. In the lists of followers/followings for a given user, those who do not have any of the desired interests are discarded. The methodology above is used to build a network of 32,900 nodes and 638,000 edges, which is used in the next simulations.
	
As discussed in Section \ref{subsection: Info Diffusion and MCS Rec}, following the IM process, influenced users register in the MCS platform, which allows gathering more information about them. For each user, information are gathered regarding their exact GPS coordinates, their average speed, the residual energy in the user’s device, and the user’s reputation. To simulate this, once the expected influence is computed using the information process, a set of MCS attributes is assigned to each influenced user. The GPS coordinates for each user are randomly generated around their general location. For example, a user located in Liverpool would be randomly given GPS coordinates within Liverpool. The user’s average speeds are assigned based on a real-life dataset of people’s mobility patterns, which includes vehicular mobility traces of users in the city of Cologne, Germany \cite{uppoor2011large,uppoor2013generation}. The residual energies are randomly generated following a uniform distribution \cite{azzam2016grs,AZZAM20181,alagha2019data}. The users’ reputations are generated using the Stack Exchange Data Dump \cite{StackExchange}. 
\vspace{-1em}
\subsection{Performance Evaluation}
To prove the efficacy of the proposed system (\textit{IIWRS}), each of its phases is compared with existing and synthetic benchmarks, via three different experiments: 

\begin{itemize}[leftmargin=*]
    \item In Section \ref{subsubsec: g-b IM simulations}, the proposed IM phase is compared against existing works, which mainly adopt individual-based greedy methods to select influencers \cite{wang2018social,lu2019task}. This comparison shows the significance of the group-based selection.
    \item In Section \ref{subsubsec: Interest-based IM experiment}, the significance of considering the interests while selecting influencers is shown. The proposed ranking metric in the IM is compared against a system with a ranking metric that only considers influencers' connections/in-degrees \cite{page1999pagerank}. 
    \item In Section \ref{subsec: full system experiment}, the complete \textit{IIWRS} system, including the recruitment stage, is compared against traditional MCS and recent OSN-MCS systems \cite{azzam2016grs,wang2018social}. 
\end{itemize}

The experiments were conducted on a Dell Intel Xeon workstation equipped with $256$ GB RAM, $300$ GB Hard Disk, and $2.6$ GHz CPU. Each data point presented is the result of averaging the final outcome of $100$ repeated simulations. Each simulation has its own IM and information diffusion process to compute the influence, after which the corresponding outcome is obtained for that particular simulation, and averaged over the $100$ simulations. For example, each point in Fig. \ref{Figs: Ug and Rg Simulations} is the result of averaging the outcome of 100 simulations, where each simulation has its own IM and information diffusion processes, and hence its own $R(g)$ or $U(g)$ outcome. The final outcome would be the average of the $R(g)$ or $U(g)$ values over the 100 simulations.

\subsubsection{Group-based IM}
\label{subsubsec: g-b IM simulations}
This experiment is conducted to validate the proposed group-based IM approach and show its superiority over the greedy individual-based one \cite{wang2018social,lu2019task}. Two ranking metrics are compared, $U^V(g)$ and $R(g)$, as defined in Section \ref{subsection: MCS&Interest IM}. In this experiment, $3$ tasks are considered, with interests randomly chosen out of the 5 aforementioned interests mentioned in Section \ref{Subsec: Dataset}.

\begin{figure}[h]
\centering
\subfloat[\label{Figs: Ug simulations}]{
\includegraphics[width=0.85\columnwidth]{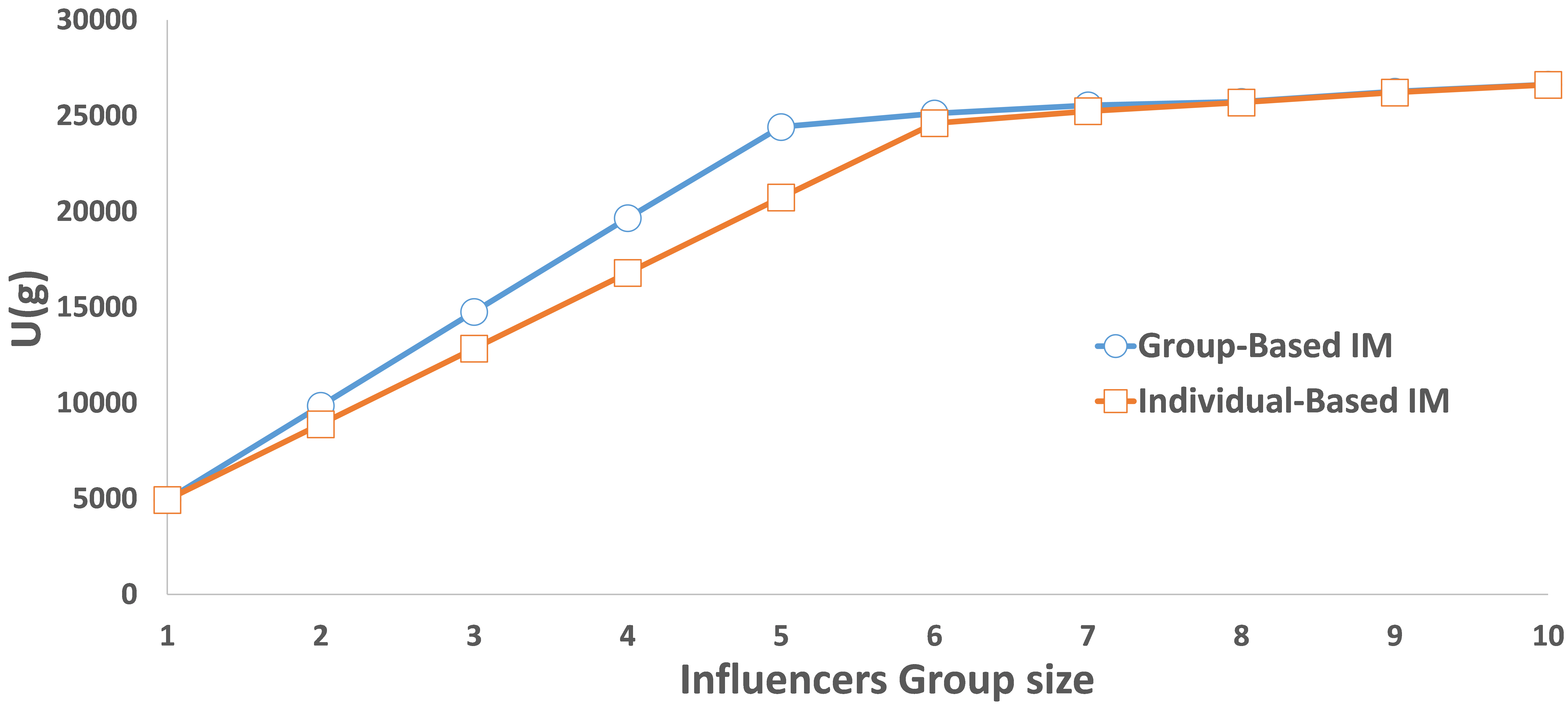}}
\hfill
\subfloat[\label{Figs: Rg simulations}]{
\includegraphics[width=0.85\columnwidth]{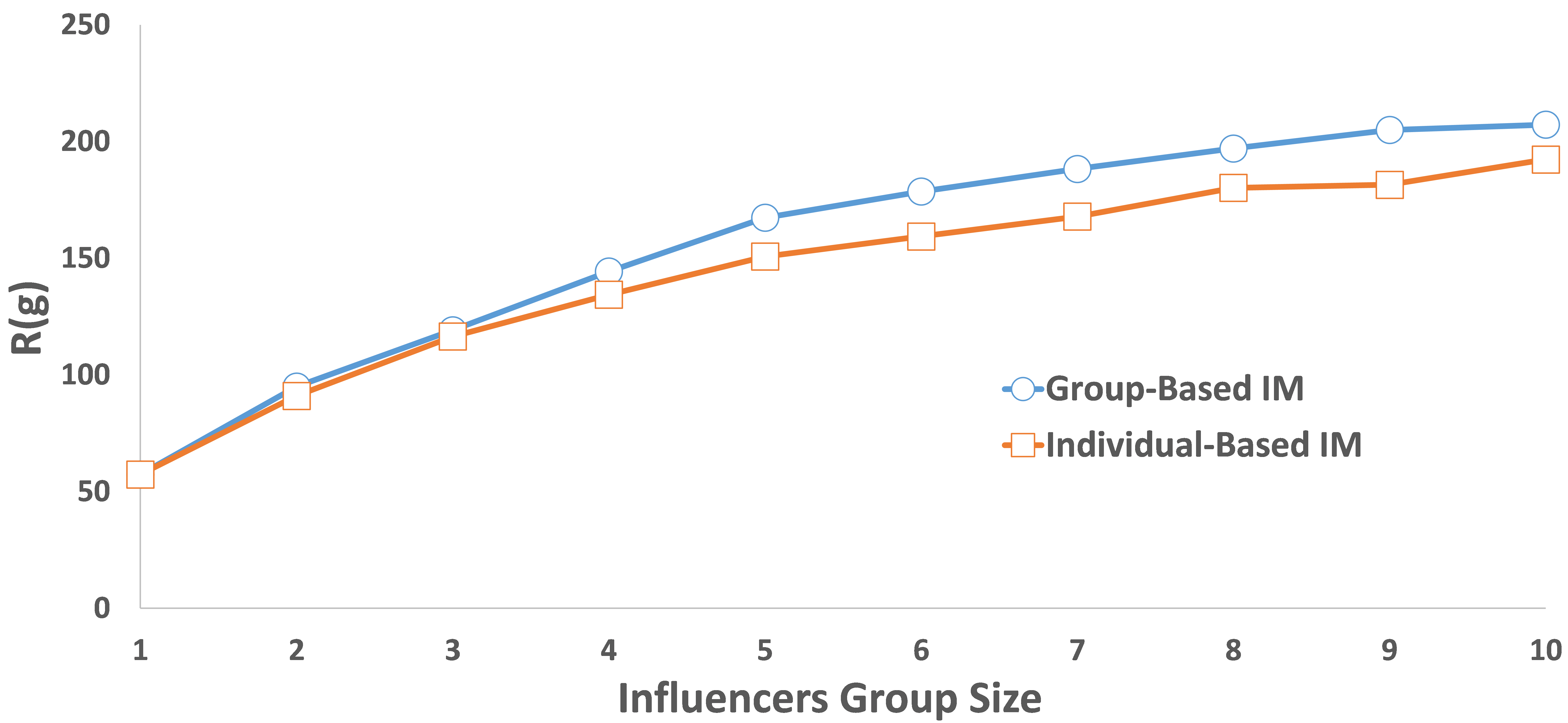}}
\hfill
\caption{Comparing the IM approaches in terms of (a) $U^V(g)$ and (b) $R(g)$.}
\label{Figs: Ug and Rg Simulations}
\vspace{-1em}
\end{figure}

Fig. \ref{Figs: Ug and Rg Simulations} evaluates this comparison, while varying the influencers group size. The group size is varied here to show that the proposed approach outperforms other methods, regardless of the group size. It can be seen that the proposed group-based approach in \textit{IIWRS} is able to provide a set of influencers with higher $U^V(g)$ and $R(g)$, when compared with the individual-based approach. Specifically, \textit{IIWRS} achieves up to $15\%$ and $12\%$ higher scores in terms of $U^V(g)$ and $R(g)$, respectively. As evident in Fig. \ref{Figs: Ug simulations}, the percentage increase of $U^V(g)$ is smaller when the group goes beyond a size of 5. The convergence of $U^V(g)$ indicates that after hiring the first 5 influencers, no additional influencers can bring significant increase to the number of followers. In Fig. \ref{Figs: Rg simulations}, the convergence occurs later, since $R(g)$ considers multiple additional parameters, and hence there is more chance for candidate influencers to contribute through any of the parameters.

\subsubsection{Interest-based IM}
\label{subsubsec: Interest-based IM experiment}
This experiment shows and validates the significance of considering users' interests when choosing influencers in \textit{IIWRS}. The proposed IM approach, which uses $R(g)$ as the ranking metric, is compared against one that uses $U^V(g)$ instead. The metric $U^V(g)$ considers only the number of followers, unlike $R(g)$ which additionally considers users' interests. For each approach, the influence of the selected set of influencers is computed using the IC diffusion model, as discussed in Section \ref{subsection: Info Diffusion and MCS Rec}, with an influence probability $p_{u,v}=0.02$. This value, obtained from \cite{statista1,statista2}, reflects the engagement rate of an “average” influencer. After computing the influence, the two approaches are compared in terms of the \textit{\textbf{interested influence}}, i.e. the subset of the influenced nodes having the desired interest(s). The setup for the rest of the experiment is the same as in Section \ref{subsubsec: g-b IM simulations}.

\begin{figure}[h]
\centering
\includegraphics[width=0.85\columnwidth]{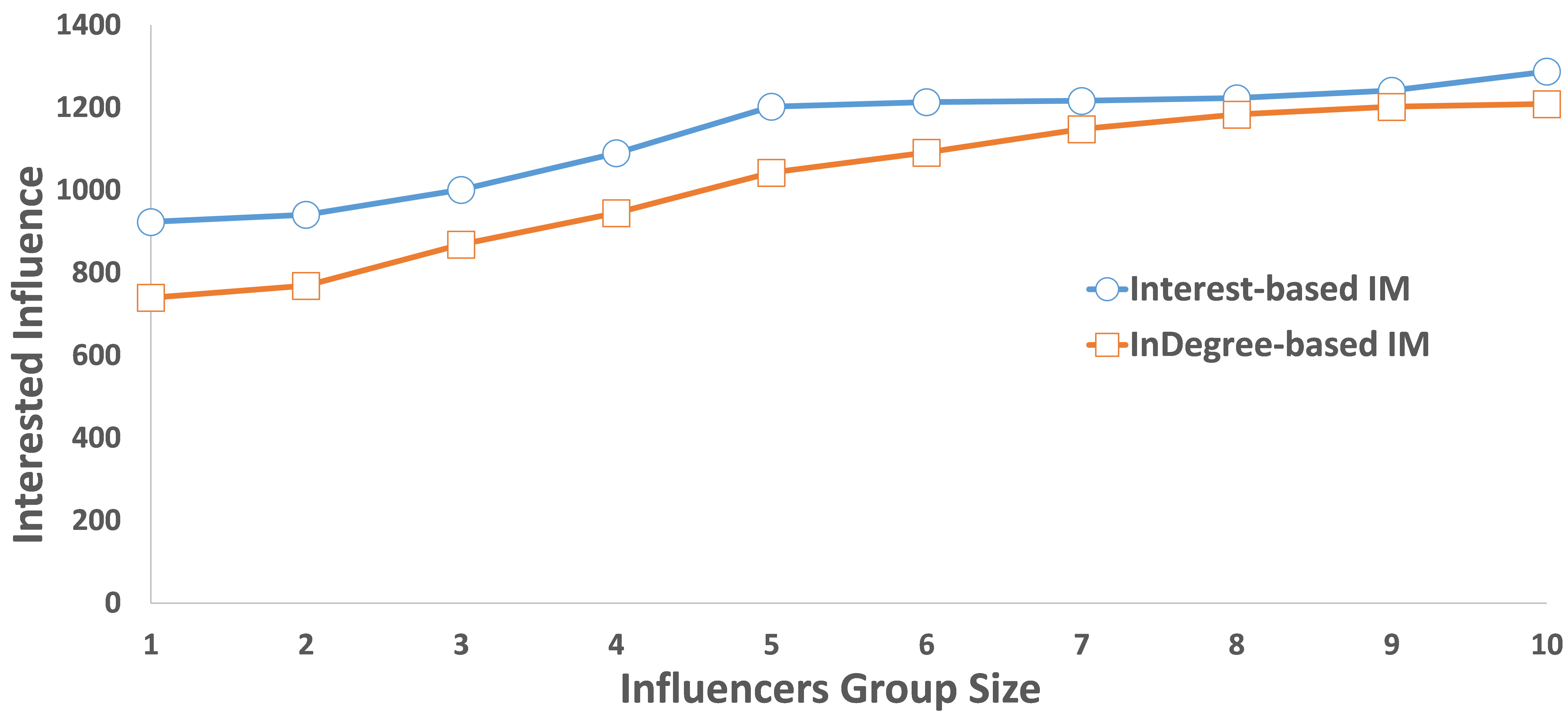}
    \caption{Comparing the two IM approaches in terms of the interested influence.} 
    \label{Figs: Interest Simulations}
    \vspace{-1em}
\end{figure}

Fig. \ref{Figs: Interest Simulations} shows the interested influence obtained using the $R(g)$ and $U^V(g)$ ranking metrics, for varying group sizes. While focusing only on the number of followers, i.e. $U(g)$, might yield higher influence, but considering the interest of influencers is shown to yield higher interested influence. An interested influence is expected to yield better quality when it comes to performing MCS tasks related to their interests. As shown in Fig. \ref{Figs: Interest Simulations}, the interest-based IM achieves higher interested influence, as high as $20\%$, when compared to the in-degree-based IM. The interested influence converges after a group size of $5$, indicating that additional influencers do not contribute much to the interested influence.

\subsubsection{Influence- and Interest-based MCS Recruitment}
\label{subsec: full system experiment}
This experiment evaluates the performance of textit{IIWRS} in comparison to the following existing and synthetic systems:
\begin{itemize}[leftmargin=*]
    \item Group Recruitment System (GRS): A group-based MCS recruitment system with a definite pool of $182$ workers \cite{azzam2016grs}. Given the pool of candidates, the aim is to maximize the QoS of the recruited group. This work does not take into account the probability of workers rejecting the task.
    \item Dynamic GRS (DGRS): A synthetic version of GRS where workers who reject the task are substituted.
    \item Social Network-Assisted Worker Recruitment System (SWRS): A system where workers are directly taken from the network \cite{wang2018social}. This work uses an individual-based IM process, as discussed in Section \ref{subsubsec: g-b IM simulations}. It is assumed in this work that influenced users are the set of recruited workers, and hence there are no considerations to the specific MCS-related attributes, such as the residual energy in their devices or their interest levels. This work does not take into account the probability of workers rejecting the task.
    \item Dynamic SWRS (DSWRS): A synthetic version of SWRS where workers who reject the task are substituted. 

\end{itemize}

A comparison between IIWRS and GRS is done to show the effect of providing a larger pool of candidate workers and the dynamic recruitment in \textit{IIWRS}, while a comparison with SWRS is done to show the significance of the proposed group-based IM and MCS recruitment stages. It is worth mentioning that the proposed dynamic recruitment process in \textit{IIWRS} is embedded in GRS and SWRS to produce the synthetic versions, DGRS and DSWRS. This is to show that, even when only a dynamic recruitment process is introduced to these systems, they are still lacking. This is due to insufficient candidate workers in DGRS, and the lack of considerations to group-based selection and MCS-related attributes in SWRS. As discussed earlier, one of the main issues in MCS is the acceptance rate of tasks by potential workers, which could be as low as $3-4\%$. Hence, the probability of a worker accepting the task is varied, and \textit{IIWRS} is compared with systems above.

\begin{figure}[h]
\centering
\includegraphics[width=0.85\columnwidth]{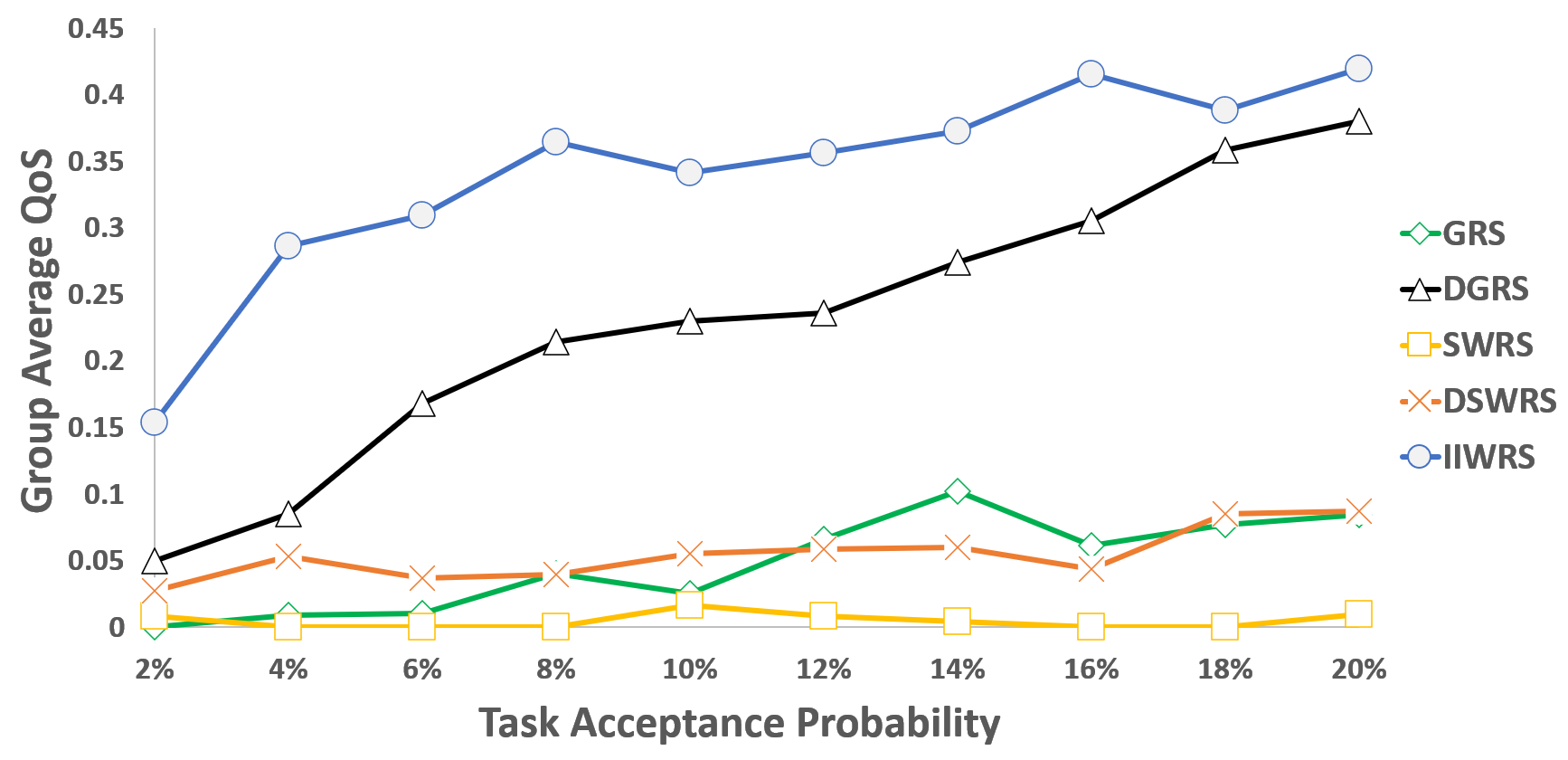}
    \caption{Comparing \textit{IIWRS} with the benchmarks in terms of the expected QoS.} 
    \label{Figs: Full System Simulations}
    \vspace{-1em}
\end{figure}

Figure \ref{Figs: Full System Simulations} compares the average QoS achieved by a recruited group of $10$ workers, for all systems, for varying percentages of task acceptance probability. The average QoS is taken over the set $QoS^W$ which contains the individual ${QoS}_j^W$ values for each group member, as computed in Equation \ref{eq: recruitment QoS}. It is worth noting that a worker $W_j$ who does not accept the task, and is not substituted, results in ${QoS}_j^W = 0$. As seen in the figure, \textit{IIWRS} is up to $30$ times better than GRS, $2.3$ times better than DGRS, $88$ times better than SWRS, and $8.5$ times better than DSWRS. \textit{IIWRS} outperforms GRS mainly due to its dynamic recruitment process, in which the workers who reject the task are substituted, which is a feature missing in GRS. Additionally, the small size of the pool in GRS makes it hard to find suitable workers. Even when dynamic recruitment is introduced in DGRS, \textit{IIWRS} still outperforms due to the large pool of candidate workers obtained from the OSN. Moreover, when compared to SWRS and DSWRS, \textit{IIWRS} shows significant improvement in the QoS. \textit{IIWRS} outperforms SWRS mainly due to 1) the group-based IM which yields higher influence as shown in Sections \ref{subsubsec: g-b IM simulations} and \ref{subsubsec: Interest-based IM experiment}, 2) the recruitment process which considers the workers MCS-related attributes obtained during the registration stage, and 3) the dynamicity of the recruitment process in substituting workers that reject the assigned tasks.
Even when the worker recruitment process is made dynamic in DSWRS, IIWRS still outperforms due to the group-based IM and the consideration to the MCS-attributes. Finally, it can be seen that even as the task acceptance probability increases, where the benchmarks' performances get better, \textit{IIWRS} still outperforms.
\vspace{-0.8em}
\vspace{-0.1em}
\section{CONCLUSION}
\label{Section: Conclusion}
In this paper, a comprehensive Influence- and Interest-based Worker Recruitment System (\textit{IIWRS}) in MCS, using OSNs, is proposed. \textit{IIWRS} uses a proxy-based influence maximization method with group-based selection to identify a set of influencers with certain desired interests to advertise about the MCS tasks. Compared to individual-based methods, this method was shown to improve the achieved ranking score by up to $15\%$, which results in selecting influencers that are more suitable to advertise about the desired tasks. It was also shown that interest considerations in the proposed IM method result in spreading out the information about the tasks to more targeted users with interests matching the tasks domain, which helps in tackling the cold start problem. Hence, an improvement of as high as $20\%$ is achieved, when compared to methods that only consider the number of followers an influencer has. After a pool of candidate workers is created through influence diffusion, a dynamic worker recruitment process, which considers worker’s social attributes, is used to recruit the best workers and ensure substitution in case workers refuse to perform the assigned tasks. The complete \textit{IIWRS} system is shown to perform up to $88$ times better than the existing benchmarks. Through these results, the proposed system shows its efficacy by identifying MCS-related influencers, reaching out to a specific and targeted group in the network that matches the desired MCS tasks, and dynamically recruiting workers and substituting those who reject the assigned task. Hence, the proposed system can overcome the practical challenges of cold start and low participation in current MCS systems.
\vspace{-1em}
\section*{Acknowledgement}
This work was supported by ADEK - Abu Dhabi Department of Education and Knowledge (AARE18-106).
\vspace{-0.8em}
\bibliography{bibliography}
\bibliographystyle{IEEEtran}

\vskip -3\baselineskip plus -1fil
\begin{IEEEbiography}[{\includegraphics[width=1in,height=1.25in,clip,keepaspectratio]{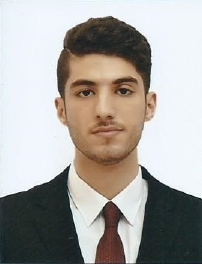}}]{Ahmed Alagha} received the B.Sc. and M.Sc. degrees in Electrical and Computer Engineering from Khalifa University, Abu Dhabi, UAE, and worked as a Research Associate in the same university. He is currently pursuing his Ph.D. at Concordia University, Montreal QC, Canada, in the Information System Engineering department (CIISE). His research interests include multi-agent systems, machine learning, artificial intelligence, IoT, sensing technologies, crowd sensing and sourcing, optimization methods, social networks, and radiation detection.
 \end{IEEEbiography}
\vskip -3\baselineskip plus -1fil
 \begin{IEEEbiography}[{\includegraphics[width=1in,height=1.25in,clip,keepaspectratio]{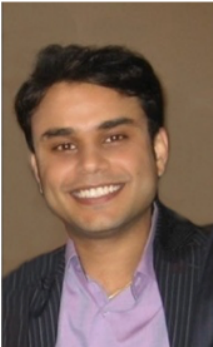}}]{Shakti Singh} received the B.Sc., M.Sc., and Ph.D. degrees in Electrical and Computer Engineering from Purdue University, West Lafayette, IN, USA. He joined the faculty at Khalifa University of Science and Technology, Abu Dhabi, UAE, in the Electrical Engineering and Computer Science department, in 2010. He pursues research related to semiconductor devices and technologies, artificial intelligence (AI), crowd sensing and sourcing, blockchain, and IoT sensors and networks.
\end{IEEEbiography}
\vskip -3\baselineskip plus -1fil
\begin{IEEEbiography}[{\includegraphics[width=1in,height=1.25in,clip,keepaspectratio]{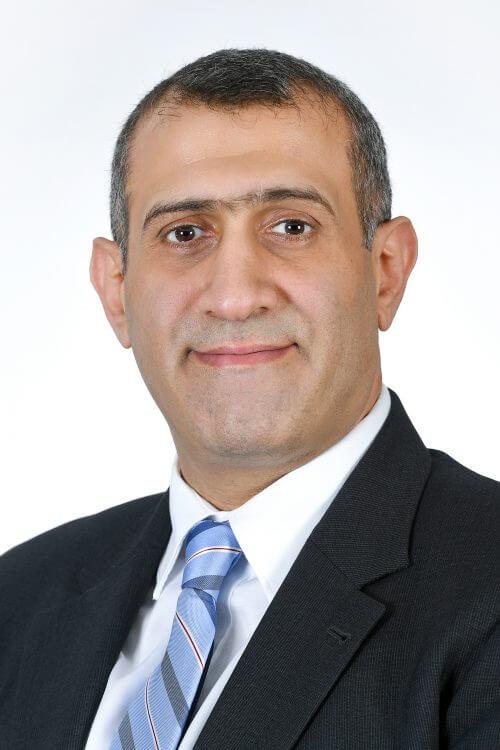}}]{Hadi Otrok} (senior member, IEEE) received his Ph.D. in ECE from Concordia University. He holds a Full professor position in the department of Electrical Engineering and Computer Science (EECS) at Khalifa University, an affiliate associate professor in the Concordia Institute for Information Systems 
Engineering at Concordia University, Montreal, Canada, and an affiliate associate professor in the electrical department at Ecole de Technologie Superieure (ETS), Montreal, Canada. He is an associate editor at: IEEE TNSM, Ad-hoc networks (Elsevier), and IEEE TSC. His research interests include: Blockchain, reinforcement learning, Federated Learning, crowd sensing and sourcing, ad hoc networks, and cloud and fog security.

\end{IEEEbiography}
\vskip -3\baselineskip plus -1fil
 \begin{IEEEbiography}[{\includegraphics[width=1in,height=1.25in,clip,keepaspectratio]{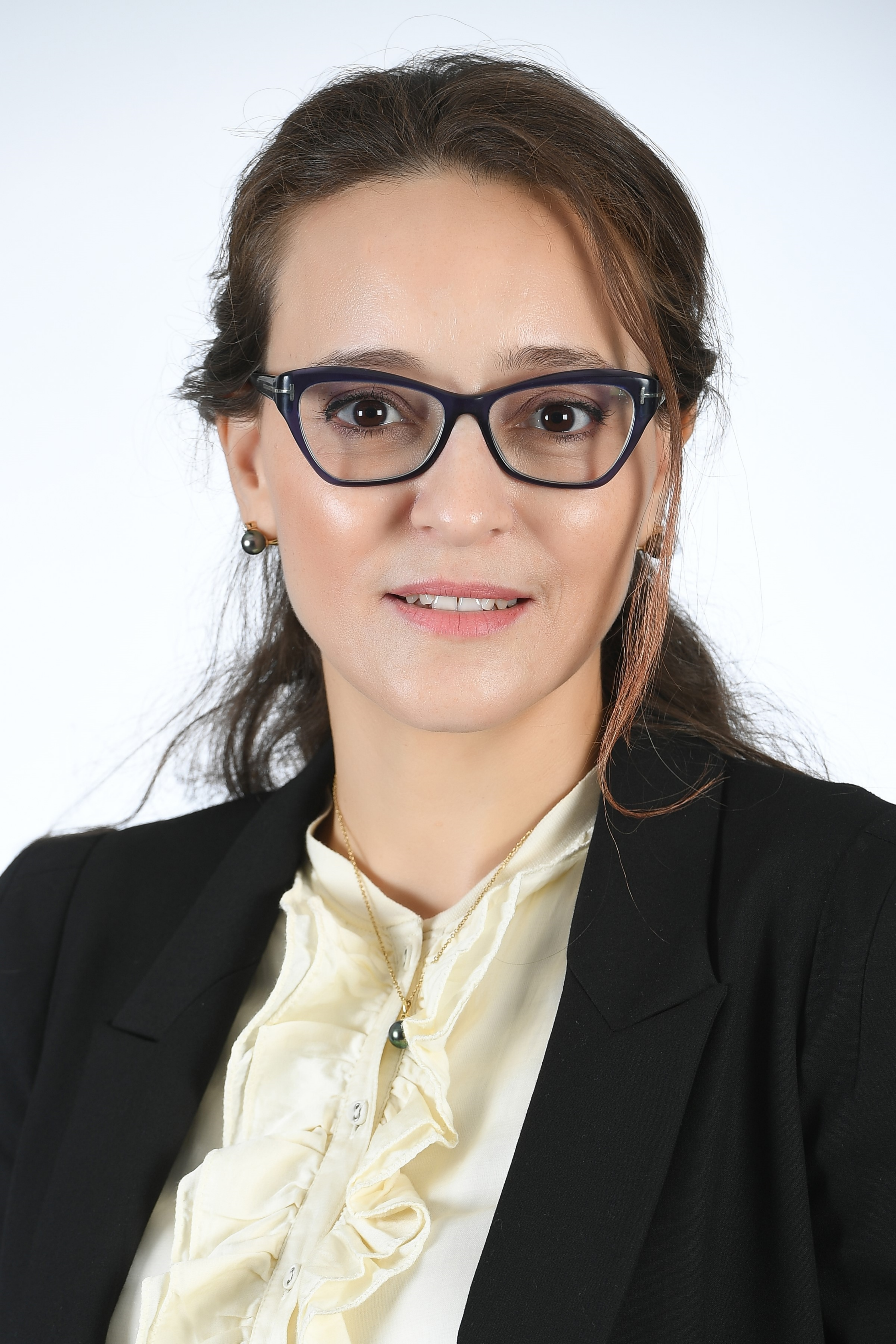}}]{Rabeb Mizouni} is an Associate Professor in the Department of Electrical Engineering and Computer Science at Khalifa University, Abu Dhabi, United Arab Emirates. She got her M.Sc. and Ph.D. in Electrical and Computer Engineering from Concordia University, Montreal, Canada in 2002 and 2007 respectively. Currently, she is interested in the deployment of context aware mobile applications, crowd sensing, software product line and cloud computing.
\end{IEEEbiography}

\end{document}